# Field emission from single multi-wall carbon nanotubes

M. Passacantando[*], F. Bussolotti, and S. Santucci
*Dipartimento di Fisica, Università degli Studi dell'Aquila and INFN and CNR-INFM Laboratorio Regionale CASTI
via Vetoio, 67010 Coppito (AQ), Italy*

A. Di Bartolomeo, F. Giubileo, L. Iemmo, and A. M. Cucolo.
*Dipartimento di Fisica E.R. Caianiello, and CNR-INFM Laboratorio Regionale SUPERMAT, and Centro
Interdipartimentale NanoMateS, Universita` degli Studi di Salerno, and INFN Gruppo Collegato di Salerno
via S. Allende, 84081 Baronissi (SA), Italy*

**ABSTRACT**

Electron field emission characteristics of individual multiwalled carbon nanotubes have been investigated by a piezoelectric nanomanipulation system operating inside a scanning electron microscopy chamber. The experimental setup ensures a high control capability on the geometric parameters of the field emission system (CNT length, diameter and anode-cathode distance). For several multiwalled carbon nanotubes, reproducible and quite stable emission current behaviour has been obtained with a dependence on the applied voltage well described by a series resistance modified Fowler-Nordheim model. A turn-on field of ~30 V/µm and a field enhancement factor of around 100 at a cathode-anode distance of the order of 1 µm have been evaluated. Finally, the effect of selective electron beam irradiation on the nanotube field emission capabilities has been extensively investigated.

[*] **Corresponding author:** Phone: +39 0862 433297; Fax: +39 0862 433033.
*e-mail address:* maurizio.passacantando@aquila.infn.it (M. Passacantando).

## I. INTRODUCTION

Since their discovery[1] the carbon nanotubes (CNTs) have attracted a considerable attention due to their peculiar and unsurpassable chemical, mechanical and electronic properties[2]. In particular, CNTs electronic transport behaviour (metallic vs. semiconducting like) may vary according to their structural conformation[2], thus suggesting the use of single walled (SWCNTs) or multi walled (MWCNTs) carbon nanotubes as conducting nanowires suitable for the investigation of the electric transport phenomena in the mesoscopic regime[3]. Within this context, due to the possibility of tuning their physical properties via a proper modification of the growing conditions (techniques and/or growing parameters), CNTs represent the ideal candidates for the development of a new class of carbon based electronic nanodevices[4,5]. In particular, the use of CNTs as field electron emitters is considered as one of the most promising technological application due to the possibility of industrial production[6-10]. Indeed, CNTs are characterized by high emission electronic properties due to their good electron conductivity and a specific geometry, resulting in a drastic amplification of the electrical field strength in the vicinity of the nanotube tip[6-10].

Up to now most of the experimental work has been devoted to the study of the field emission (FE) properties of CNT dense films as grown on metallic and semiconducting substrates[6-10]. In particular, the electron emitting properties of the CNT films result to be deeply affected by the overall organisation of the nanotubes (alignment degree, density, length, etc.) and on the quality and typology of the supporting material. In the case of a densely packed film, however, the number of CNTs involved in the electronic emission process is hard to estimate, with the additional complication that the emission capabilities of individual sites can vary widely, due to the dependence of the CNTs electronic properties on structural parameters as diameter, chirality, occurrence of structural defects, etc[1,2]. At the same time, the importance of adsorbates in determining the current enhancement and the saturation current during the field emission process has been clearly brought to light[10]. Such a lack of an adequate control capability on the electronic, structural and chemical properties of the nanotube's films makes rather problematic a detailed modelling of the field emission process of CNT films. For the same reasons, it is difficult to draw any quantitative conclusions about the field emission properties of individual CNTs starting from measurements carried on nanotube films. Consequently, experiments performed on individual nanotubes are valuable in providing a fundamental understanding of the emission mechanism, representing a simple and structurally controlled system to be immediately compared with the theoretical models of the field emission process. Moreover, the importance of investigating the behaviour of individual CNTs as field emitters is also related to their technological potentialities as nano-sized electron sources in developing high resolution electron beam based instrumentation as displays[11] and electron microscopes[12].

Field emission from a single MWCNT has been recently reported by several authors[13-15], who have measured emission currents up to several tens of µA with different techniques. It has been shown that long time emission usually results in a drastic CNT structure damage, leading to sudden emission failures. It has also been proven that both the length ($\ell$) and the radius (*r*) of the tube play an important role in the FE process, influencing both the field enhancement factor γ and the turn-on field $E_t$. These latter two parameters have been extensively studied as a function of the CNT-anode separation (d): an asymptotic behaviour characterized by a value independent from *d* has been found at the "far field condition", i.e. for $d > 3\,\ell$, respectively after an increasing (for γ) and a decreasing (for $E_t$) phase.

In this paper, the *in situ* field emission characterization of individual MWCNTs by means of a nanomanipulation system operating inside a Scanning Electron Microscope (SEM) is presented. The experimental setup ensures a high control on the field emission geometric parameters (CNT length and diameter, cathode-anode shape and separation)[16]. Reproducible emission current behaviour up to densities of $10^5$ A/cm$^2$ has been measured, with a dependence on the applied voltage well described by a series resistance modified Fowler-Nordheim model over more than 7 orders of magnitude. An estimation of the turn-on field as low as 30 V/µm has been determined, with a field enhancement factor around 100 at a CNT-anode distance of ~1 µm. Furthermore, as remarkable result, the effect of a selective electron beam irradiation on the nanotube field emission capabilities has been extensively investigated, showing the possibility of switching off the FE process by causing a modification of the CNT apex.

## II. EXPERIMENTAL

A two MM3A-nanoprobe system manufactured by Kleindeik Company has been used for the CNT manipulation and the electrical transport measurements. The system has been installed inside a SEM (ZEISS, LEO 1430). Multiwalled carbon nanotubes have been directly grown onto a nickel wire by means of a Chemical Vapour Deposition (CVD) technique[16]. The nickel wire has been first polished in acetone by a 30 minutes sonication cycle in order to remove all the surface contaminants, then it has been inserted in a CVD reactor, pumped down to less than $10^{-6}$ Torr using a turbo molecular pump. The specimen has been annealed at 700 °C for 5 minutes in NH$_3$ gas at a flow rate of 100 sccm/min. The CNTs have been grown on the metallic wire by adding C$_2$H$_2$ at a flow rate of 20 sccm for 20 min at the same temperature of the NH$_3$ annealing treatment, with the MWCNT growth catalysed by the protrusions and clusters of the nickel surface[16]. During both the annealing and the CNTs growth, the pressure inside the quartz reactor has been kept steady at 5.5

Torr. The resulting CNTs are distributed quite uniformly on the metallic wire surface, sometimes organized in bundles from which free-standing nanotubes protrudes (Fig. 1(a)).

Both manipulators were coupled with a Keithley 4200-SCS, operating as source-measurement unit. The CNT covered nickel-wire (anode) has been fixed on one of the nanomanipulators. The opposite counter-electrode was constituted by an electrochemically etched tungsten tip (cathode), connected to the second nanomanipulator that can be easily brought into contact with individual nanotubes of the nickel wire for manipulations and field emission measurements. All the field emission measurements have been performed in a high vacuum atmosphere ($<10^{-6}$ Torr) of the SEM chamber at room temperature. In order to better reproduce the ideal two-parallel plate electrode configuration, the initially sharp tungsten tip (curvature radius ~100 nm, Fig. 1(a)) has been intentionally rounded up to ~4 μm of curvature radius (Fig. 1(b)) *via* cathode-anode arc discharge production[17].

Individual MWCNTs have been attached onto the cathode rounded tungsten probe and current measurements have been carried out with an applied voltage varying from 0 to a maximum of 120 V, the limit voltage allowed by the nanomanipulators circuits. The emitted current was measured with a sensitivity better than 0.1 pA. The equivalent circuit is modelled in Fig. 1(c), where *R* is a series resistance accounting for the CNT, the interfaces, the contacts, the wires and the measurement devices resistances. A rough estimation of the last three contributions has been determined by bringing the cathode (before the rounding process) into direct electrical contact with the anode, thus measuring an overall resistance value of about 12 kΩ.

The rounded tip has been carefully moved into contact with individual protruding multiwalled nanotubes, the electrostatic interaction favouring a spontaneous adhesion of the nanotube onto the metal surface. Once a first and quite stable CNT/metal contact has been obtained, a high current value ($10^{-4}$ A) has been applied across the entire tungsten-CNT-metal system, thus resulting, via Joule heating effect, in the cutting of the CNT in correspondence of a defective point[18,19]. At the same time, the current produced an anneal of the tungsten-nanotube interface, due to heating stimulated contaminants desorption[18,19], yielding the formation of stable, though quite resistive, metal-nanotube contact. Furthermore, the open and sharp CNT tip morphology, which is a direct consequence of the current induced cutting procedure, is expected to enhance the emitting properties of the CNT, making the resulting system particularly suitable for field emission investigation[5].

According to preliminary investigations, a current limit of 1 μA has been imposed during the data acquisition, in order to avoid CNT damages induced by the current heating, as severance or complete destruction of the emitting structure. At the same time this limit enables reproducible measurements with respect to the applications of several voltage sweeping cycles.

## III. RESULTS AND DISCUSSION

Figure 2(a) shows a SEM image of an as-obtained suspended 6 μm long CNT, having a diameter of about 60 nm (Fig. 2(b)), with a CNT tip-Nickel anode distance of about 1.0 ± 0.4 μm. The high error on the distance estimation is due to the difficulty to project the nanotube tip on the highly rough opposite surface. The corresponding field emission characteristic curves (as acquired in two consecutive sweeping voltage) have been reported in Fig. 2(c). An emission threshold field, defined here as the applied field required for a current of 1 pA, can be evaluated as $E_t = V/d = 35 \pm 14$ V/μm.

Field emission results are usually analysed in the framework of the Fowler-Nordheim (F-N) theory, which describes the emission of electrons from a metal as a quantum-mechanical tunnelling enhanced by a strong electric field[20]. The Fowler-Nordheim equation can be written as:

$$I = S \cdot a \frac{E_s^2}{\Phi} \exp\left(-b \frac{\Phi^{3/2}}{E_s}\right) \quad (1)$$

where $S$ is the emitting surface, $E_s$ is the electric field at nanotube apex, $a$ and $b$ are constants (when $S$, $\Phi$ and $E_s$ are expressed in cm$^2$, eV and V/cm, respectively, the constants $a$ and $b$ assume the values of $1.54 \times 10^{-6}$ AeVV$^{-2}$ and $6.83 \times 10^7$ VcmeV$^{-2/3}$, respectively). The electric field at the nanotube tip $E_s$ is related to the applied voltage $V$ via $E_s = \gamma V/d = \gamma E_t$, being $\gamma$ the field enhancement factor at the sharp tip of the nanotube and $d$ the CNT tip-anode separation. According to the F-N model, a plot of $\ln(I/V^2)$ vs. $1/V$ (known as F-N plot) has a linear behaviour with a slope that can be used to evaluate the field enhancement factor. In particular, as clearly shown in Fig. 2(d), the F-N plots for the MWCNT in Fig. 2(a) and 2(b) can be easily fitted by straight line (dashed line) over a rather wide range of electric fields, a result that dispels the need to replace the F-N model with a novel emission mechanism. Assuming a CNT work function of 4.8 eV [Ref. 21], for the CNT of Fig. 2 a value of $\gamma = b\Phi^{3/2}d/m = 63 \pm 25$ has been determined, in agreement with those obtained in different experimental studies[13-15].

The slight deviation from the linear behaviour as observed in the high voltage region of the F-N plot (Fig. 2(d)) or, equivalently, the saturation in the log($I$) vs. $V$ curve (Fig. 2(c)) has been easily explained by taking into account the additional voltage drop across the series resistance $R$ (see Fig. 1(c)). A fit of the F-N data by using a properly modified F-N equation, i.e.

$$I = c_1(V - RI)^2 \exp\left(-\frac{c_2}{V - RI}\right) \quad (2)$$

(where $c_1$, $c_2$ and $R$ are the fitting parameters) is also reported in Fig. 2(d) (continuous line). A good accordance with the experimental data, on the entire measurement range, is obtained by a value $R \sim$ 30 MΩ. This high resistance value, in part due to nanotube heating and defects, is believed to be prevalently contributed by the contact resistance at the metal-CNT interface.

The time stability of the FE current has been investigated by using the experimental geometry described in Fig. 3(a), where the SEM image of a suspended 8 μm long MWCNT (diameter 80 nm, see inset of Fig. 3(a)), at a distance of about 2.2 μm from the nickel counter electrode is reported. Current fluctuations with amplitude of the order of 100% of the intensity are observed (Fig. 3(b)). Nevertheless a steady behaviour is achieved and kept for about 35 minutes before a drastic drop of the current (see inset of Fig. 3(b)), likely due to a CNT severing; then, after about 3 hours of low emission, the current slowly turns off with a gradual CNT degradation till its complete disappearance.

A more accurate estimation of the threshold emission field has been carried out without using the absolute CNT–anode distance $d$, which can be affected by a high systematic error, but relaying only on the precise and controllable relative distance as given by the nanomanipulator-driven probe movements. In particular by inverting the role of cathode and anode, with a CNT (length ~1.1 μm, diameter ~50 nm, see Fig. 4(a) and corresponding inset) on the Ni wire emitting toward the rounded tungsten probe, several *I-V* curves have been registered by stepping backward the tungsten probe, with movements of 200–300 nm, on which a 10% error is safely assumed (Fig. 4(b)). The plot of the threshold voltage as a function of the distance, as reported in Fig. 4(c), shows a linear behaviour and confirms that the local electric field required on the CNT for the emission is not a function of $d$ over the spanned range. This result can be explained by the fact that, as far as the probe is close enough to the tip of the CNT ($d \ll 3\ell$), the probe-to-CNT system approximates a parallel plate configuration, with constant $E_t$ and modest field enhancement factor. Consequently, the slope of the straight line fitting the $V_t$ vs. $d$ plot, $\partial V_t/\partial d$, can be used to evaluate the threshold field, i.e. $E_t \approx (31 \pm 4)$ V/μm. This differential method has the advantage of leaving apart the initial probe-CNT separation, and by using only the precise nanomanipulator movements[22]. Finally, the F-N plots of each emitting curves have been analysed in the framework of linear F-N model (Fig. 4(d)) thus evaluating the correspondent field enhancement factor γ. The plot of γ as a function of CNT tip-anode distance, as reported in Fig. 4(e), demonstrates the tendency of the field enhancement factor to rise with the distance, a behaviour which has been reported by several authors both for single CNT emitters[14,15] and in vertically aligned CNT films[22].

To complete the FE characterization of single MWCNTs, we also investigated the innovative capability to perturb (switching off) the emission process by modifying the tube apex by electron beam irradiation. To this aim the field emission geometry reported in Fig. 5(a) was considered, where the SEM image of a suspended CNT (length ~4 μm, diameter ~70 nm) is shown[1]. In particular a 10 keV electron beam has been focused on the CNT terminating region (80 nm × 80 nm of irradiation area). After 10 min of exposure a clear morphological modification has been observed, with an overall smoothing of the initially sharp, open ended CNT apical region (see inset of Fig. 5(a) and 5(b) for comparison) resulting, at the same time, in a complete suppression of the field emission currents, as reported in Fig. 5(c), even in correspondence, in the case of exposed CNT, of higher voltage values.

As clearly emerges by the comparison of Fig. 5(a) and 5(b) the use of the combined nanomanipulator-SEM system ensures a complete stability of the geometry (i.e. CNT-cathode distance) of the field emission experiment before and after the electron irradiation cycle. Therefore, the dramatic drop of the CNT emitting capability can be univocally ascribed to the morphological modification of the CNT apical region, as induced by the electron beam exposure. In this context, the SEM electron beam (10 keV) is not expected to produce any deep structural modifications of the CNT irradiated area (C atom knock out, C vacancy, etc) as observed in the case of high energetic (>80 keV) electrons bombardment[23]. On the other hand, the low damaging interaction between the CNT and the 10 keV electron beam has been clearly demonstrated by independent electronic transport measurements[16]. The shape modification of the nanotube apex can be more easily interpreted in term of electron induced deposition of carbonaceous. The pressure inside the SEM chamber, in fact is $6 \times 10^{-6}$ Torr, with a small fraction of residual hydrocarbon molecules naturally present in the system. The hydrocarbon molecules decompose under the electron beam irradiation, thus resulting in the formation of a carbonaceous deposit onto the irradiated area[16,19,24]. In particular, the carbonaceous deposit has been demonstrated to be in the form of amorphous carbon which acts as an insulator with a resistivity of about $10^{11}$ Ω cm [Ref. 19, 25]. The electron induced deposition of an insulating carbonaceous on the CNT apical region layer therefore introduces an additional barrier at the CNT/vacuum interface which can be associated with the observed suppression of the field emission current. By comparing the SEM images of the CNT apical region before and after the irradiation a rough estimation of the main geometric parameter (i.e. thickness and section area) of the amorphous carbon layer can be performed. In particular (see Fig. 2(b) and 2(c) for comparison) after the irradiation the carbonaceous material deposit can be roughly assumed to be distributed in a cylindrical volume of 40 nm thickness and having the same section area of the supporting CNT (diameter ~60 nm). The curve with black circle of Fig. 5(c), showing no significant current between the irradiated CNT and the tungsten anode, demonstrates the crucial role of the amorphous cap in suppressing the field emission capability of the CNT. Such an high efficiency combined with the high selectivity of the SEM beam focusing process opens the route to a more systematic study of the field emission properties of CNT large arrays and to the possibility of modulating the current by controlling the performances of each single emitter. In particular by means of electron beam induced deposition of carbonaceous materials the controlled and selective suppression of the field emission from every single CNT is made possible, allowing detailed study of the screening effect on the field emission capabilities of CNT bundles or vertically aligned films.

---

1 The CNT in Fig. 5 results from the breaking of the CNT in Fig. 2, as obtained by current induced heating.

**IV. CONCLUSIONS**

The results of in situ field emission measurements from single MWCNT inside a high vacuum SEM chamber has been reported. In particular, the behaviour of the field emission current as a function of the applied voltage and the time has been studied in a wide current range and with high sensitivity. The key field emission parameters as the threshold field and the field enhancement factor (about 30 V/µm and 100, respectively) have been evaluated. We have shown that the emission phenomenon is very well described by a series resistance modified Fowler-Nordheim model.

Finally an original method to limit or suppress the field emission current from a single nanotube through the controlled and selective deposition of carbonaceous species on its apical region by means of a low energy electron focused beam has been presented. This result opens several perspectives for the study of the field emission properties of CNT arrays, enabling the switching of selected single emitters, as well as for technological applications where tunable nano-sized emitters might be needed.

**ACKNOWLEDGMENTS**

This work has been partially supported by Ministero dell'Istruzione, dell'Università e della Ricerca Scientifica (PRIN n. 2005020415).

**FIGURE CAPTIONS**

**Figure 1.** SEM images of the tungsten tip before (a) and after (b) intentional rounding by arc discharge. (c) Equivalent circuit of the field emission experimental setup. The overall resistance R takes into account the contributions of the CNT, the contact resistance at the metal/CNT interface, the connection wires and the Source Measurement Unit.

**Figure 2.** (a) SEM images representing the geometry of the field emission experiment performed on a suspended CNT. (b) Detail of the CNT apical region, acquired at higher SEM magnification. (c) CNT field emission curves, as obtained in two consecutive voltage sweeps. (d) F-N plot of the field emission data of panel (c). The F-N data of the two voltage sweeps have been plotted together (black dot). The linear (blue line) and resistance modified (red line) F-N fittings are also reported (see text for details).

**Figure 3.** (a) SEM image of ~8 μm long suspended MWCNT. A higher SEM magnification of the CNT apical region is reported in the inset. (b) FE evolution as a function of time. The current has been measured at a fixed bias of 45 V.

**Figure 4.** (a) SEM image of a ~1.1 μm-long CNT (diameter ~ 50 nm, see inset) emitting electron towards a rounded tungsten probe. (b) *I-V* characteristic and threshold voltage at different probe-CNT tip distances. For each curve the threshold voltage (defined in correspondence of an emission current of 1 μA - see dashed line) has been evaluated. (c) Threshold voltage values as a function of the CNT tip-probe distance (black circles). According to the linear interpolation (solid line) a field enhancement factor of (31 ± 4) V/μm has been estimated. (d) F-N plots (dot) and corresponding linear interpolation (solid line) of the *I-V* data reported in panel (a). (e) Field enhancement factor plotted as a function of the CNT tip-anode distance.

**Figure 5.** (a) SEM image representing the geometry of the field emission experiment, performed on a suspended CNT. A closer SEM image of the CNT apical structure is reported as inset. (b) SEM image of the field emission geometry after a prolonged electron beam exposure of the tip region (reported as inset). (c) *I-V* characteristic of the CNT before (white square) and after (black circle) the electron beam exposure.

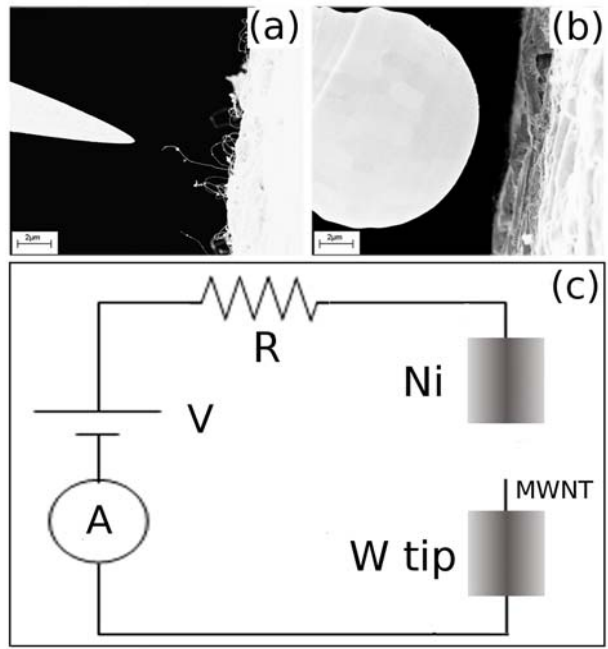

**Figure 1**

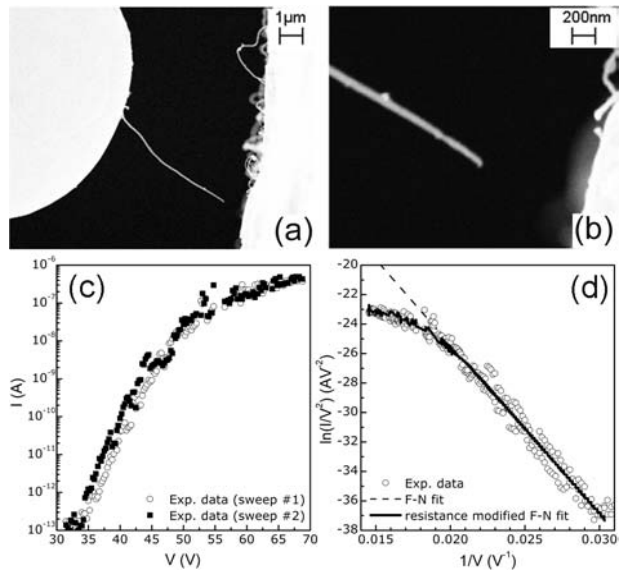

**Figure 2**

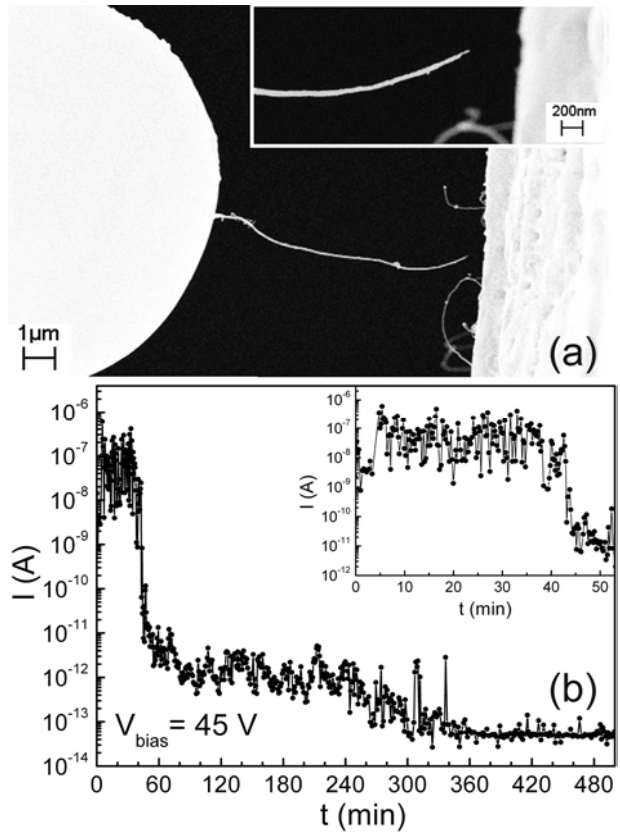

**Figure 3**

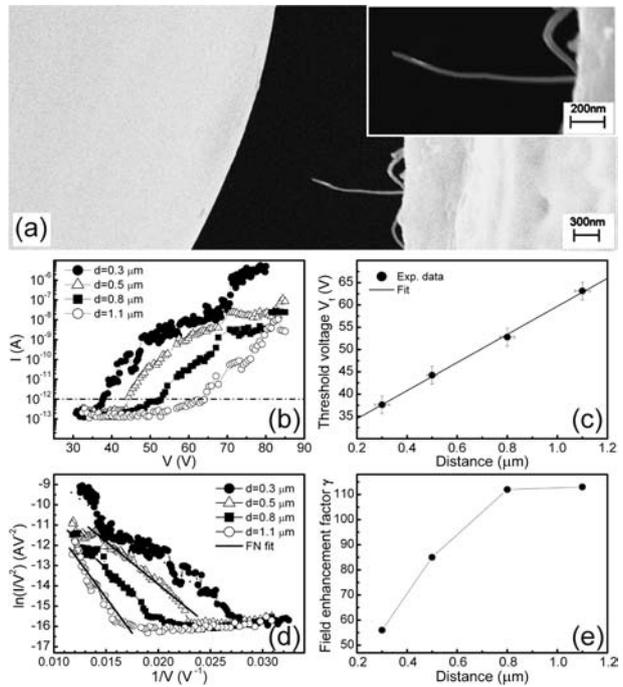

**Figure 4**

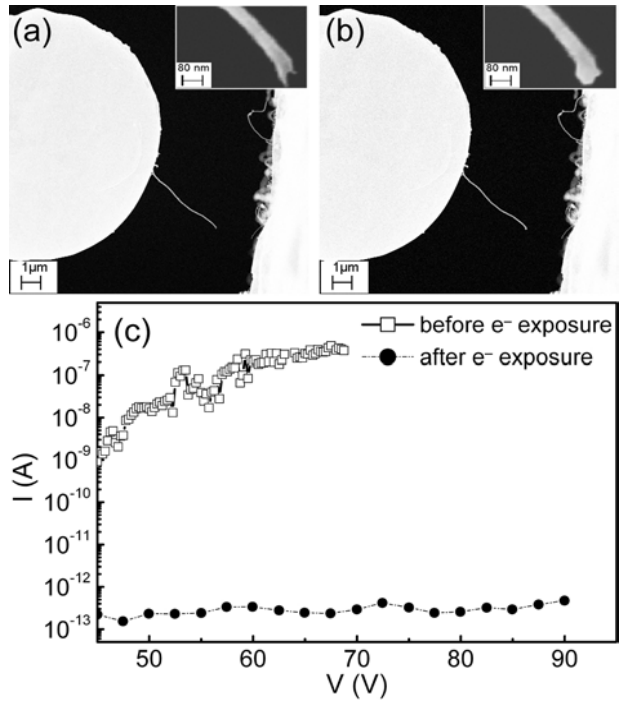

**Figure 5**